\documentclass[twoside]{LCWS11}
\usepackage[latin1]{inputenc}
\usepackage[dvips]{graphicx,epsfig,color}
\usepackage{wrapfig,rotating}
\usepackage{amssymb,amsmath,array}
\usepackage{cite}
\begin{document}
\title{
%%%%   Paper title goes here  %%%%%%%%%%%%%%
Hadron Energy Resolution of the CALICE AHCAL
 and Software Compensation Approaches} %% 
%***********************************************************************
% AUTHORS INFORMATION AREA
%***********************************************************************
\author{Marina Chadeeva on behalf of the CALICE Collaboration
% Optional short acknowledgment: remove next line if non-needed
%\thanks{
%This is an optional funding source acknowledgment.
%}
% DO NOT MODIFY THE FOLLOWING '\vspace' ARGUMENT
\vspace{.3cm}\\
% Addresses and institutions (remove "1- " in case of a single institution)
Institute for Theoretical and Experimental Physics - Moscow, Russia
%1- First Author's Institution - Department \\
%Address of First Author's Institution - Country
%% Remove the next three lines in case of a single institution
}
%%***********************************************************************
% END OF AUTHORS INFORMATION AREA
%***********************************************************************

\maketitle

\begin{abstract}
The hadron energy resolution of a highly granular CALICE analogue scintillator-steel hadronic calorimeter was studied using pion test beam data. The stochastic term contribution to the energy resolution was estimated to be $\sim$58\%$/\sqrt{E/\mathrm{GeV}}$. To improve an energy resolution, local and global software compensation techniques were developed which exploit an unprecedented granularity of the calorimeter and are  based on event-by-event analysis of the energy density spectra. The application of either local or global software compensation technique results in reducing of stochastic term contribution down to $\sim$45\%$/\sqrt{E/\mathrm{GeV}}$. The achieved improvement of single particle energy resolution for pions is $\sim$20\% in the energy range from 10 to 80~GeV. 
\end{abstract}

\section{Introduction}

To evaluate detector technologies for the future linear collider experiments~\cite{ILC} the CALICE collaboration~\cite{CALICE}  has constructed and commissioned high-granular electromagnetic and hadronic calorimeter prototypes that have been successfully operated in test beam experiments at DESY, CERN and Fermilab since 2006 till 2011. During the test beam campaigns a large amount of data was accumulated using muon, electron and pion beams in the energy range from 1 to 180~GeV with different detector configurations. The  unprecedented granularity of the CALICE calorimeter prototypes allows to analyze hadronic shower structure with high spatial resolution~\cite{EcalHadrons} and test particle flow algorithms using test beam data~\cite{PandoraTest}.

The hadronic energy resolution of the CALICE analogue scintillator-steel hadronic calorimeter was studied using test beam data. We also present a brief description of developed software compensation techniques as well as the results of their successful application to improve a single particle energy resolution for pions.

\section{Experimental setup and data}

\subsection{CALICE test beam setup}
The data collected during 2007 test beam campaign at CERN SPS with positive and negative pion beams in the energy range from 10 to 80~GeV were analyzed. During this data taking period, CALICE test beam setup consisted of  silicon-tungsten electromagnetic calorimeter (ECAL)~\cite{ECAL:2008}, scintillator-steel analogue hadronic calorimeter (AHCAL)~\cite{AHCAL:2010cc} and scintillator-steel tail catcher and muon tracker (TCMT)~\cite{TCMT}. The test beam setup was also equipped with various trigger and beam monitoring devices including \u{C}erenkov counter.

The Si-W ECAL is a sampling calorimeter with a total depth of 24 radiation lengths. In this study the ECAL was used for event selection and early shower detection. Events with a primary inelastic interaction in the ECAL were rejected.

The AHCAL consists of small 5-mm thick plastic scintillator tiles with individual readout by silicon photomultipliers (SiPMs). The tiles are assembled in 38 layers (each layer 90$\times$90~cm$^2$) interleaved by 2-cm thick stainless steel absorber plates. The size of the scintillator tiles ranges from 3$\times$3~cm$^2$ in the central region,  6$\times$6~cm$^2$ in the outer region and 12$\times$12~cm$^2$ along the perimeter of each layer. In the last eight layers only 6$\times$6~cm$^2$ and 12$\times$12~cm$^2$ tiles  are used. In total, the CALICE AHCAL has 7608 scintillator cells and amounts to a depth of 4.5 nuclear interaction lengths.

The TCMT consists of 16 readout layers assembled from scintillator strips read out by SiPMs between steel absorber plates and has two sections with different sampling fractions. In this study the information from TCMT was used for muon separation and to minimize the effect of leakage from the AHCAL at higher energies.

\subsection{Calibration and event selection}

The calorimeter calibration procedures are described in detail in~\cite{ECAL:2008,AHCAL:2011em}. The visible signal in cells produced by a minimum-ionizing particle (MIP) was studied with muons and is taken as a base unit of visible energy measurement. To reject noise, only signals above the threshold of 0.5~MIP are used in the analysis, hereinafter these signals are called hits. 

The total deposited energy is obtained at the ``electromagnetic scale''. The conversion factors from the visible signal in units of MIP to the total energy in units of GeV are extracted from electromagnetic calibration of each subdetector. The conversion factor from visible signal to deposited energy in the ECAL for non-showering hadrons was estimated using simulated muons and the measured muon response from test beam data sets.  For the conversion to the hadronic energy scale, the $\frac{e}{\pi}$ factor was considered in addition to calculate the initial reconstructed energy, as discussed below. 

The event selection procedure provides a purification of the pion samples that have an admixture of muons as well as electrons or protons. The efficiency of muon identification was checked using both simulated muon samples and muon samples from test beam data and was estimated to be better than 98\%.  The muon identification results in a purity of pion samples from muons better than 0.5\% in the studied energy range. The \u{C}erenkov counter is used to remove electrons from the $\pi^{-}$ samples and protons from the $\pi^{+}$ samples. 

To analyze the intrinsic AHCAL energy resolution for hadrons, an additional constraint is applied to the purified pion samples: a shower start (a position of the primary inelastic interaction) was required to be in the first five layers of the AHCAL. This requirement allows the effect of leakage into the TCMT to be reduced by selecting hadronic showers which are mostly contained in the AHCAL.

\subsection{Hadron energy reconstruction and intrinsic energy resolution}

Without software compensation the initial reconstructed energy $E_{\mathrm{initial}}$ of an event is calculated as follows:
\begin{equation}
E_{\mathrm{initial}} = E_{\mathrm{ECAL}}^{\mathrm{track}} + \frac{e}{\pi} \cdot \left(\sum_{\mathrm{hit}} E_{\mathrm{hit}} + E_{\mathrm{TCMT}}\right),
\label{eq:Eini}
\end{equation}
\noindent where $E_{\mathrm{ECAL}}^{\mathrm{track}}$ is the energy lost by a hadron track in the ECAL, $\frac{e}{\pi} = 1.19$ is a scaling factor to take into account the different response to electrons and hadrons in the non-compensating AHCAL (the coefficient was obtained for pions by averaging the ratio of beam energy to the total energy reconstructed at electromagnetic scale over the studied energy range), $E_{\mathrm{hit}}$ is the energy deposited in one AHCAL cell and $E_{\mathrm{TCMT}}$ is the energy deposited in the TCMT.

%\begin{figure}[!h]
\begin{wrapfigure}{r}{0.95\columnwidth}
{\includegraphics[width=0.31\columnwidth]{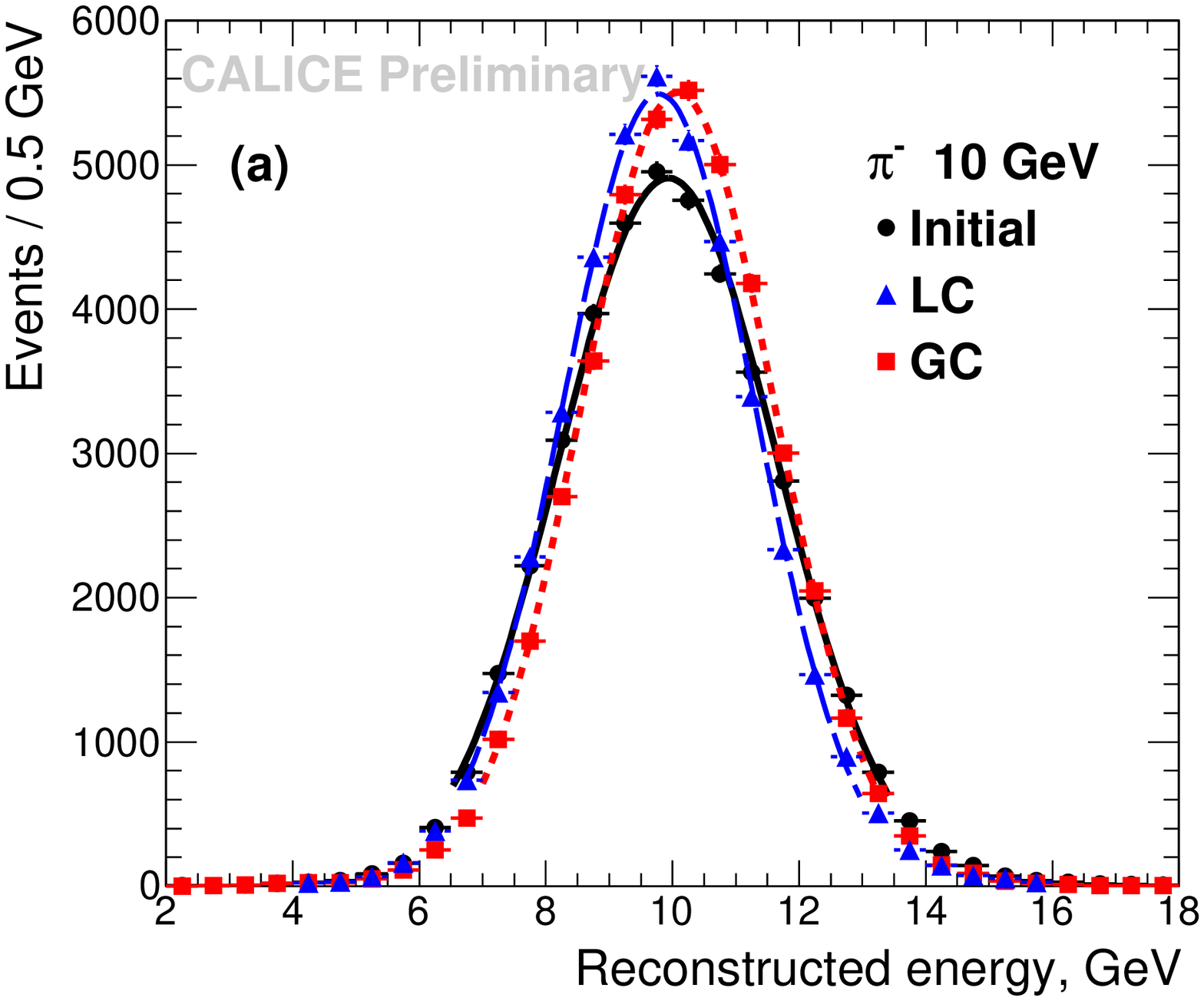}}
{\includegraphics[width=0.31\columnwidth]{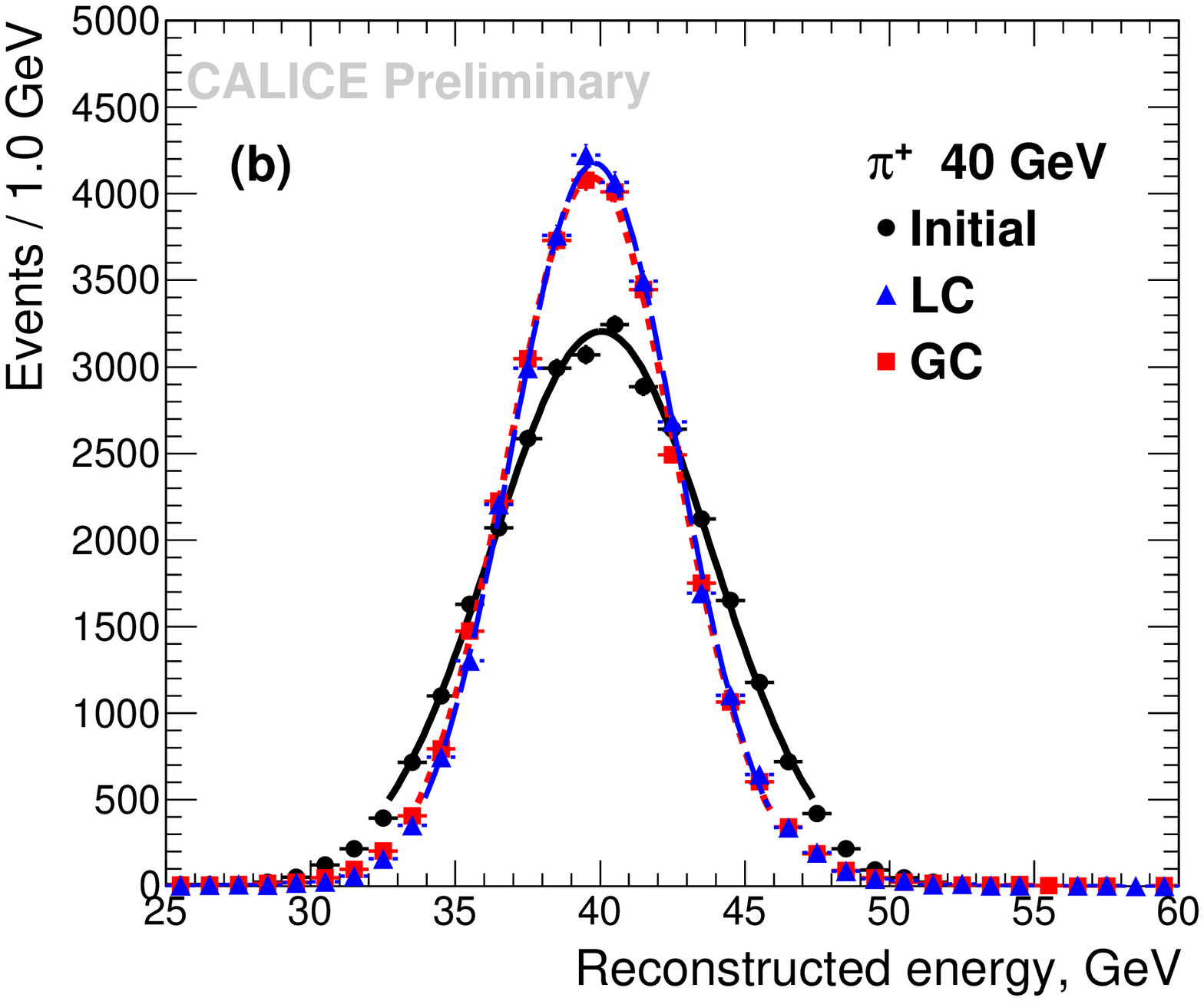}}
{\includegraphics[width=0.31\columnwidth]{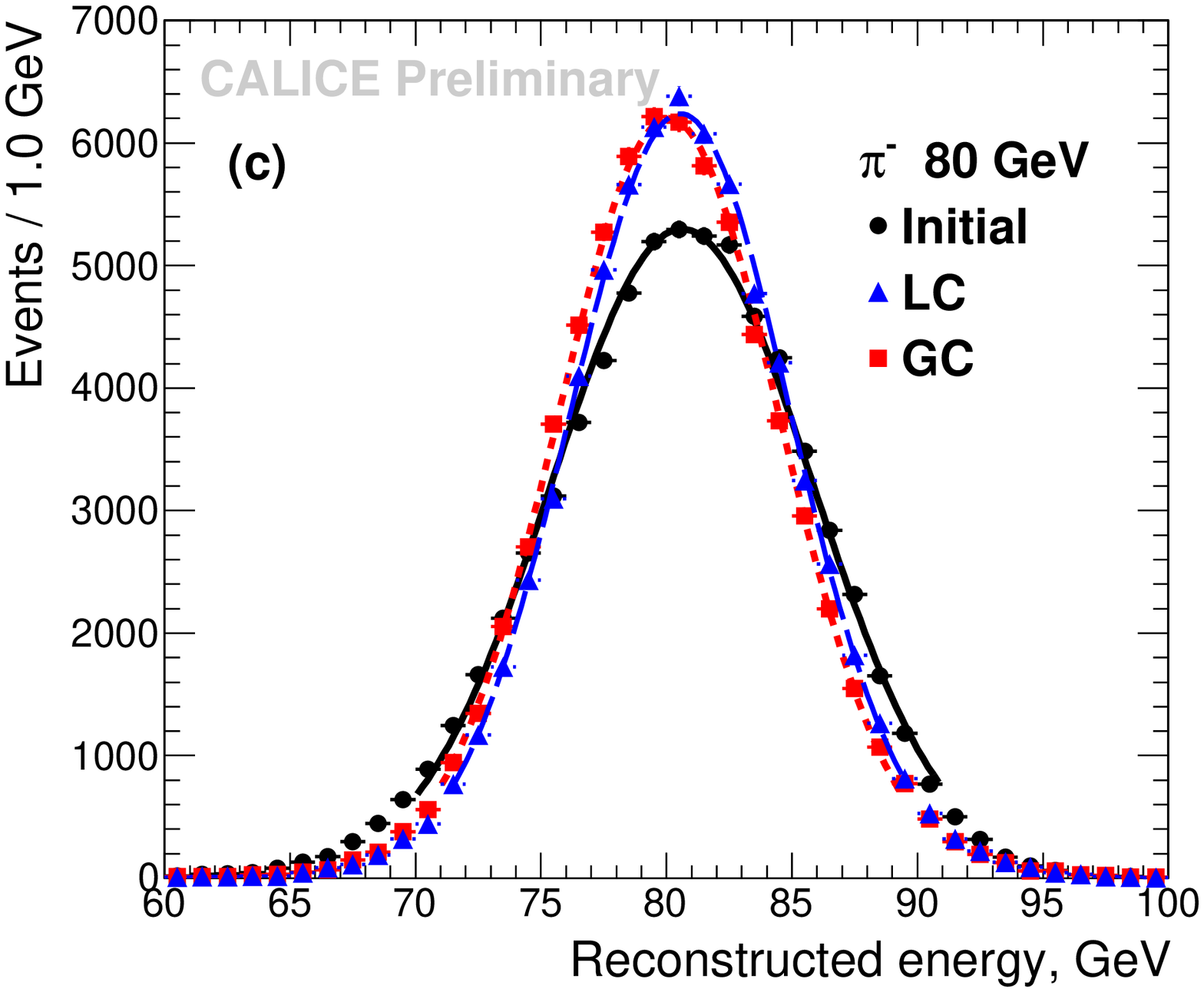}}
\caption{Reconstructed energy distributions for pions at 10~GeV (a), 40~GeV (b) and 80~GeV (c) without compensation (black circles and solid line) and after local (LC - blue triangles and dashed line) and global (GC - red squares and dotted line) software compensation applied. The curves show a Gaussian fit to the corresponding distributions. Statistical errors are shown.}
\label{fig:enrDist}
\end{wrapfigure}
%\end{figure}

\begin{wrapfigure}{r}{0.52\columnwidth}
%\begin{figure}
\centerline{\includegraphics[width=0.45\columnwidth]{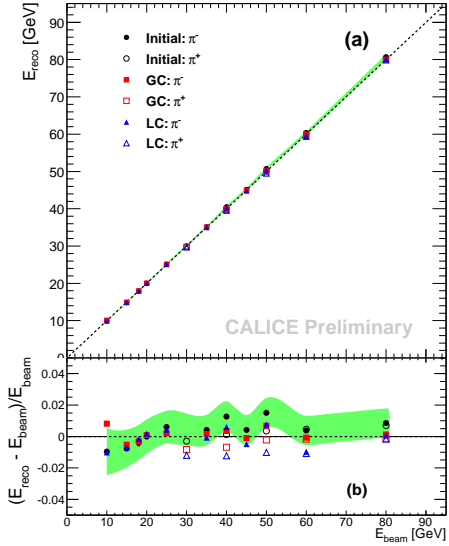}}
\caption{(a) Linearity of response to pions and (b) relative residuals to beam energy versus beam energy without compensation (black circles) and after local (LC - blue triangles) and global (GC - red squares) compensation. Filled and open markers indicate $\pi^{-}$ and $\pi^{+}$, respectively. Dotted lines correspond to $E_{\mathrm{reco}} = E_{\mathrm{beam}}$ and green band shows systematic uncertainties for the initial $\pi^{-}$ data sample. }
\label{fig:lin}
\end{wrapfigure}
%\end{figure}

\vspace{0.5cm}
The reconstructed energy distributions were fitted with a Gaussian in the interval of $\pm 2$ RMS around the mean value. Hereinafter, the mean and sigma of this Gaussian fit at a given beam energy are referred as a mean reconstructed energy $E_{\mathrm{reco}}$ and resolution $\sigma_{\mathrm{reco}}$, respectively. Fig.~\ref{fig:enrDist} shows three examples of the energy distributions for three different beam energies. The initial reconstructed energy distribution is shown by the black circles and its Gaussian fit is indicated by the black solid curve.

The response of the calorimeter setup to pions as a function of beam energy is shown in Fig.~\ref{fig:lin}a, where black circles correspond to the mean initial reconstructed energy without software compensation. Due to the intrinsic non-compensation of the CALICE AHCAL, the response to pions is non-linear with energy, deviating $\pm$2\% from a perfectly linear behavior in the studied energy range. 
In Fig.~\ref{fig:lin}b the relative residuals to the true beam energy are shown. The green band in Fig.~\ref{fig:lin} corresponds to the systematic uncertainty of the reconstructed energy with the main contribution from the uncertainty of MIP to GeV conversion coefficient for the AHCAL which  was extracted from electromagnetic calibration with an accuracy of 0.9\% \cite{AHCAL:2011em}.

\begin{wrapfigure}{r}{0.5\columnwidth}
%\begin{figure}
\centerline{\includegraphics[width=0.45\columnwidth]{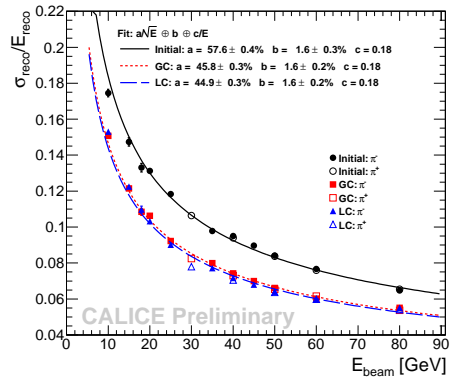}}
\caption{Relative resolution versus beam energy without compensation (black circles and solid line) and after local (LC - blue triangles and dashed line) and global (GC - red squares and dotted line) compensation. The curves show fits using eq.~{\protect \ref{eq:relres}}. }
\label{fig:res}
\end{wrapfigure}
%\end{figure}

The relative energy resolution is shown in Fig.~\ref{fig:res}, where black circles correspond to the initial energy resolution without software compensation. The resolution for $\pi^{-}$ events (filled markers) is in good agreement with that observed for $\pi^{+}$ events (open markers). The solid curve represents a fit with the following function:
\begin{equation}
 \label{eq:relres}
 \frac{\sigma}{E} = \frac{a}{\sqrt{E}} \oplus b \oplus \frac{c}{E},
\end{equation}
\noindent where $E$ is in GeV, and $a$, $b$ and  $c$ are stochastic, constant and noise contributions, respectively. The noise term is fixed at $c = 0.18$~GeV, corresponding to the measured noise contribution in the full CALICE setup taking in account contributions from ECAL (0.004~GeV), AHCAL (0.06~GeV) and TCMT (0.17~GeV). These estimates were obtained using dedicated runs without beam particles as well as using random trigger events constantly recorded during data taking \cite{AHCAL:2011em}.

The stochastic term contribution to the initial hadron energy resolution of the AHCAL was estimated to be $\sim \frac{58\%}{\sqrt{E/\mathrm{GeV}}}$. The constant term contribution is $\sim 1.6\%$. The similar relative energy resolution was observed for $\pi^{+}$ and $\pi^{-}$ samples. The application of software compensation methods described below allows to decrease the contribution from the stochastic term and to improve the hadron energy resolution for the AHCAL.

\section{Software compensation}

A response of sampling calorimeter to hadrons is much more complicated than to electrons and comprises the contributions from two different components: electromagnetic (mostly from $\pi^0$ particles and nuclear photons) and hadronic. The latter includes a so called ``invisible'' component (neutrons, nuclear binding energy losses, etc.) that remains undetectable leading to lower calorimeter response to hadrons than to electrons. The fractional containment of the mentioned components fluctuates significantly from event to event \cite{Wigmans:2000,Groom:2007}.  The existence of the ``invisible'' component and event-by-event fluctuations of electromagnetic and hadronic fractions inside hadronic showers are two factors that deteriorate a hadronic energy resolution compared to electromagnetic one.

For intrinsically non-compensating calorimeters, compensation can be achieved by using an approach known as ``off-line'' or ``software'' compensation and based on the expectation of higher energy density inside electromagnetic component compared to a hadronic one. Such procedures have been successfully implemented for several calorimeters, e.g. WA1 calorimeter~\cite{Abramovitz:1981}, LAr calorimeter of H1 experiment~\cite{H1:2005} and ATLAS hadronic endcap calorimeter~\cite{ATLAS:2004}.

To improve the energy resolution, two software compensation techniques based on hit spectrum analysis were developed for the AHCAL which allow to decrease the contribution from stochastic term. Both approaches exploit a high granularity of the CALICE hadronic calorimeter and are based on event-by-event  analysis of the energy density spectrum obtained from the signals of individual calorimeter cells. At the same time the approaches follow different ways to construct and apply compensation factors: in the local approach there are several weights applied to the signals of individual calorimeter cells, while in the global approach one compensation factor is applied to the energy sum calculated over all calorimeter. 

\subsection{Local software compensation technique}

In the local software compensation procedure different weights are applied to every cell of the AHCAL. The values of these weights reflect whether a cell belongs to a hadronic or electromagnetic sub-shower. To distinguish between different types of energy deposit on the cell level, the local energy density in each particular cell is analyzed. 
%Fig.~\ref{fig:local} shows an example of the energy density spectrum divided by bins depending on the energy density.
As the cells in electromagnetic sub-showers typically have a higher energy deposit, they get a lower weight in the overall energy sum compare to cells with a lower energy density. The energy of the event corrected by the local software compensation technique can be calculated as follows:
\begin{equation}
E_{\mathrm{LC}} = E_{\mathrm{ECAL}}^{\mathrm{track}} + \sum_{\mathrm{hit}} E_{\mathrm{hit}} \cdot \omega_{\mathrm{hit}} (E_{\mathrm{initial}}) + \frac{e}{\pi} \cdot E_{\mathrm{TCMT}},
\label{eq:eventELC}
\end{equation}
\noindent where $\omega_{\mathrm{hit}}(E_{\mathrm{initial}})$ is a hit weight which depends on the hit energy density and on the total initial event energy. The procedure of adjusting weights and extracting parameters of their energy dependence is described in more detail in \cite{CaliceNote15}. 

\subsection{Global software compensation technique}

The global software compensation procedure uses one observable that describes the shape of the hit energy spectrum for a given event.  For a higher electromagnetic fraction one can expect on average a higher energy deposition per calorimeter cell and therefore a larger relative contribution of high energy hits to the event hit spectrum. For each event the global compensation factor
 $C_{\mathrm{gl}}$ is calculated as a ratio of the number of shower hits with measured visible signal in MIP below threshold $e_{\mathrm{lim}}$ to the number of shower hits with measured visible signal below the mean hit energy in the given spectrum. The threshold $e_{\mathrm{lim}}$ is set to 5~MIP. 
The energy of the event corrected by the global software compensation technique can be calculated as follows:
\begin{equation}
E_{\mathrm{GC}} = E_{\mathrm{ECAL}}^{\mathrm{track}} + P(C_{\mathrm{gl}},E_{\mathrm{initial}}) \cdot \left( \sum_{\mathrm{hit}} E_{\mathrm{hit}} + E_{\mathrm{TCMT}} \right),
\label{eq:eventEGC}
\end{equation}
\noindent where $P(C_{\mathrm{gl}},E_{\mathrm{initial}})$ is a polynomial function to take into account the energy dependence of the compensation factor. Three parameters of this polynomial are extracted from test beam data. The detailed description of the global compensation technique is presented in \cite{CaliceNote28}.

\subsection{Application of software compensation to test beam data}

Both software compensation techniques described above do not require a prior knowledge of particle energy for the compensation to be applied. To take into account the energy dependence of compensation factors and weights, initial reconstructed energy is used for the calculation of corrected energy on event-by-event basis. The parameters of the energy dependence of compensation factors or weights are derived using test beam data. To assure the sample independence of the methods, the selected test beam data set was split into two subsets: the parameters for both techniques were adjusted using one subset and then applied to another subset. It should be emphasized that the application of either local or global compensation approach does not distort a Gaussian form of the initial energy distributions.

In fig.~\ref{fig:lin} the response to pions is shown after applying local or global compensation technique comparing to initial response without compensation. In general, the response remains linear after compensation within $\pm$1.5\% that is slightly better than before compensation.

The relative energy resolution before and after compensation is shown in fig.~\ref{fig:res}. A good agreement between the  $\pi^{-}$ and  $\pi^{+}$ samples was observed. The application of software compensation results in a decrease of the stochastic term contribution while the constant terms before and after compensation agree within errors. Both compensation techniques show very similar performance, with the local software compensation providing a slightly smaller stochastic term.

\begin{wrapfigure}{r}{0.5\columnwidth}
%\begin{figure}
\centerline{\includegraphics[width=0.45\columnwidth]{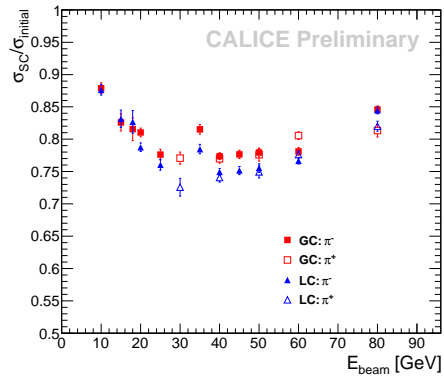}}
\caption{ Energy dependence of the relative improvement of resolution for local (LC - blue triangles) and global (GC - red squares) software compensation techniques.}
\label{fig:ratio}
\end{wrapfigure}
%\end{figure}

Fig.~\ref{fig:ratio} shows a relative improvement of the energy resolution achieved by both developed techniques.  
We define the relative improvement as the ratio of the resolution after software compensation $\sigma_{\mathrm{SC}}$ (local or global) to the initial resolution $\sigma_{\mathrm{initial}}$. As follows from Fig.~\ref{fig:ratio}, the improvement due to software compensation for test beam data ranges from $\sim$12\% to $\sim$25\% in the studied energy range, the local approach demonstrating 3\% higher improvement from 30 to 60~GeV. The worsening of the relative improvement at higher energies can be caused by a distortion of hit energy spectra due to increasing leakage into the TCMT. In the local compensation procedure, the shower hits from the TCMT are not weighted. In the global compensation procedure, the partial absence of hits in the hit spectrum results in wrong (or shifted) estimate of the global compensation factor.

\section{Conclusion}

The hadronic energy resolution of the high-granular CALICE hadronic calorimeter was studied using test beam data collected in 2007 on CERN SPS. The calorimeter comprises 7608 cells in 1~m$^3$ and is highly segmented in both longitudinal and transverse direction.  The intrinsic relative energy resolution of the CALICE AHCAL for pions is $\sim \frac{58\%}{\sqrt{E/\mathrm{GeV}}}$. The contribution of the constant term was estimated to be about 1.6\%. The unprecedented granularity of the  CALICE AHCAL allows to apply a software compensation approach based on hit spectrum analysis to improve a single particle energy resolution. Two techniques were developed and tested on data samples. The local software compensation technique applies different weights to individual cells while the global technique uses one compensation factor applied to the sum of cell signals. The achieved relative improvement of hadronic energy resolution varies from 12\% to 25\% in the studied energy range resulting in the reduction of the stochastic term contribution down to $\sim \frac{45\%}{\sqrt{E/\mathrm{GeV}}}$. Both local and global software compensation techniques give almost similar gain in resolution.

The proposed techniques do not require a knowledge of particle energy for the software compensation to be applied, the initial reconstructed energy is used as a first approximation to calculate compensation factors and weights, which are energy dependent. Although this energy dependence places some restrictions on the implementation of both proposed approaches in jet reconstruction procedure, their application in the frame of particle flow algorithms is possible when disentangled single particle clusters are considered.

\section{Acknowledgments}

The author would like to thank Frank Simon and Katja Seidel for providing their results concerning developing and application of the local software compensation technique. The author is also very grateful to Vasiliy Morgunov, Oleg Markin, Mikhail Danilov, Erika Garutti, Sergey Morozov, Angela-Isabela Lucaci-Timoce and Nils Feege for the fruitful discussions as well as to David Ward for his very helpful and important comments. 

% ****************************************************************************
% BIBLIOGRAPHY AREA
% ****************************************************************************

\begin{footnotesize}
% IF YOU DO NOT USE BIBTEX, USE THE FOLLOWING SAMPLE SCHEME FOR THE REFERENCES
% ----------------------------------------------------------------------------

% ----------------------------------------------------------------------------

\end{footnotesize}

% ****************************************************************************
% END OF BIBLIOGRAPHY AREA
% ****************************************************************************

\end{document}